
\headline={\ifnum\pageno=1\firstheadline\else
\ifodd\pageno\rightheadline \else\leftheadline\fi\fi}
\def\firstheadline{\hfil}
\def\rightheadline{\hfil}
\def\leftheadline{\hfil}
        \footline={\ifnum\pageno=1\firstfootline\else\otherfootline\fi}
\def\firstfootline{\rm\hss\folio\hss}
\def\otherfootline{\hfil}

\font\tenrm=cmr10
\font\tenit=cmti10
\font\elevenbf=cmbx10 scaled\magstep 1
\font\elevenrm=cmr10 scaled\magstep 1
\font\elevenit=cmti10 scaled\magstep 1

\font\ninerm=cmr9

\nopagenumbers
\line{\hfil }
\vglue 1cm
\hsize=6.0truein
\vsize=8.5truein
\parindent=3pc
\baselineskip=10pt
\vglue 12pt
\centerline{\elevenbf TOTAL AND JET PHOTOPRODUCTION CROSS SECTIONS}
\vglue 0.2cm
\centerline{\elevenbf AT HERA AND FERMILAB\footnote{*}
{\ninerm\baselineskip=11pt  Talk presented at the International
Conference
on Elastic and Diffractive Scattering
(Vth Blois Workshop), Providence, RI, June 8-12, 1993.
\hfil}}
\vglue 5pt
\vglue 1.0cm
\centerline{\elevenrm Ina Sarcevic }
\baselineskip=13pt
\centerline{\tenit Physics Department, University of Arizona}
\baselineskip=12pt
\centerline{\tenit Tucson, Arizona 85721}

\vglue 0.8cm
\centerline{\tenrm ABSTRACT}
\vglue 0.3cm
{\rightskip=3pc
 \leftskip=3pc
 \tenrm\baselineskip=12pt
 \noindent
 We present results of calculations of the total and jet
photoproduction cross sections at
 HERA and Fermilab energies$^1$.
 The calculations take
into account the high-energy QCD structure of the photon and are performed
for different photon  structure functions. We discuss how
recent measurements of the total
photoproduction cross section at HERA energies$^2$
can
provide important information on the low $x$ behavior of the
photon structure function
and in a more general context,
to the nature of strong interactions.
In addition, we show that
the photoproduction cross section measurements
at Fermilab E683 energies
could provide a firmer value for $p_{T}^{min}$, the lower
bound on the transverse momentum of outgoing jets, which signals the onset
of hard scattering.
The extrapolation of
our cross section to ultra-high energies, of
relevance to the cosmic ray physics, gives significant
contribution to the ``conventional'' value, but
cannot account for the anomalous muon content
observed in the cosmic ray air-showers associated
with astrophysical point sources$^3$.
\vglue 0.8cm}
\elevenrm
\line{\elevenbf 1. Introduction \hfil}
\bigskip
\baselineskip=14pt

Recent photoproduction measurements at HERA$^2$ have provided
important confirmation of the hadronlike
character of the
high-energy photon.
The probability of the photon
producing a
$q \bar q$ pair, and then through
 subsequent QCD evolution filling up the confinement volume with
quarks and gluons with a density akin to that of a pion,
increases with
energy.  Thus, it is not surprising that
even low-energy measurements of the total photoproduction cross section
up to $\sqrt s=18GeV$$^4$
show some rise with energy
similar to the one
observed in low-energy hadronic collisions$^5$.
In hadronic collisions, the rapid growth
of the total cross section
was associated with a
dominance of hard scattering partonic processes
over the nonperturbative (soft) ones$^6$, supported by the
detection
of the semi-hard QCD jets (so-called
``minijets'') at CERN Collider energies$^7$.
Similarly, recent
observation of the
hard scatterings in photoproduction at HERA energies
corroborates this
hypothesis$^8$.
\bigskip
\line{\elevenbf 2. Photoproduction Cross Sections at HERA and Fermilab
Energies
\hfil}
\vglue 0.4cm
The total photoproduction cross
section measured in the low-energy range $10GeV\leq \sqrt s\leq 18GeV$$^4$
 and, most recently, at HERA energies$^2$ points
towards
the hadronic behavior of the photon.
Few years ago,
we have made predictions for the total and jet
photoproduction cross sections
in a simple QCD minijet-type model based on analogy with
hadronic collisions$^1$ .
We have assumed that
the total photoproduction cross section
can be represented as
a sum of the soft (nonperturbative) and hard (jet) part
(i.e. $\sigma_{T} = \sigma_{soft} + \sigma_{JET} $), where
the soft part is energy
independent, determined
from the low-
energy
data ($\sqrt s\leq 10GeV$).
The jet (hard) part has contributions from
two subprocesses:
the ``standard'' (direct) QCD process ($\gamma q\rightarrow q g$ and
$\gamma g\rightarrow q\bar q$) and the ``anomalous'' process (for
example, $\gamma\rightarrow q\bar q$, followed by quark
bremsstrahlung,
$q\rightarrow qg$ and $gg\rightarrow gg$).
The later process is the same as the jet production process in
p-p collisions up to a photon structure function.
We note that the photon structure function is
proportional to
$\alpha_{em}/\alpha_{s}$, where $\alpha_{em}$ is the
 electromagnetic
coupling. The effective order of the above processes is therefore
  $ \alpha_{em}\alpha_{s},
$ since the jet cross sections are of order $\alpha_{s}^{2}$.
Thus, they
 are  of
the same order as direct  two-jet processes, in which the
photon-parton
vertex is electromagnetic and does not involve the photon's
hadronic content.
The
existing parametrizations of the photon structure function,
Duke and Owens (DO), Drees and Grassie (DG) and Abramowicz et
al. (LAC1)$^9$,
all describe
low-energy photoproduction
data very well.  However, they differ dramatically at
very high energies (i.e. low $x$ region), the region of
the HERA experiment, for example.  Therefore
recent photoproduction
measurements at HERA energies can provide valuable information
about the photon structure function at
small $x$ ($x\sim 4p_t^2/s$), and in particular its gluon
content.
\par
The QCD jet cross section for
photon-proton interactions is given by
$$
\sigma_{\rm QCD}^{\gamma p}=
 \sum\limits_{ij}{1\over{1 + \delta_{ij}}}
  \int dx_{\gamma}dx_{p}
 \int_{p_{\perp,\rm {min}}^{2}}
  \hskip -20pt dp_{\perp}^2
  [f_{i}^{(\gamma)}(x_{\gamma},\hat{Q}^{2})f_{j}^{(p)}
  (x_{p},\hat{Q}^{2}) + i \leftrightarrow j]
 {d \hat{\sigma}_{ij} \over {dp_{\perp}^2}},
  \eqno(1) $$
\noindent where $\hat {\sigma_{ij}}$ are
parton cross sections and
$f_{i}^{(\gamma)} (x_{\gamma}, \hat{Q}^{2})$
$(f_{j}^{(p)}(x_{p},
\hat{Q}^{2}))$ is the photon
(proton) structure function.
The expressions for all the subprocesses that contribute to
$\hat {\sigma_{ij}}$ can be found in Ref. 10.
We take the choice
of scale $\hat {Q^2} = p_T^2$, which is shown to give very good
description of the hadronic jet data$^{10}$.
For parton
structure function we use EHLQ parametrization$^{11}$.
The results do not show appreciately sensitivity to
the choice of the proton structure function.

{}From the constant low-energy data$^4$, we determine
the soft part of the cross section to be
$\sigma_{soft}=0.114mb$.  The observed $3\%$
increase of
the cross section in the energy range between $10GeV$ and $18GeV$
can be described by adding the
hard (jet) contribution
with jet transverse
momentum cutoff $1.4GeV\leq p_{\perp,\rm {min}}\leq 2GeV$
to the soft part$^1$.  The actual value of
$p_{\perp,\rm {min}}$, however, below which nonperturbative
processes make important contributions, is impossible to pin
down theoretically using perturbative techniques.
As the energy increases direct and soft part become
negligible in comparison with the anomalous part,
because the later has a steep increase with energy.
We find that
in the Fermilab E683 energy range ($\sqrt s\leq 28GeV$), the
results for the cross sections are not sensitive to the choice
of the photon structure function.  Therefore,
in addition to providing important confirmation of the
hadronlike nature of photon-proton interactions,
one could use
forthcoming E683 experiment to pin down the theoretical
parameter $p_{\perp,\rm {min}}$ to a few percent (see Fig. 2 in
Ref. 1).
\par
In Fig. 1 we show our results for
$\sigma_{\gamma p}$ at HERA
energies$^1$ .  We note that the results are very sensitive to the
choice of the photon structure function due to their
different $x$ behavior at very high energies. For example,
DO gluon structure function behaves as $f_g^{\gamma}
= {x^{-1.97}}$,
while DG has less singular behavior, $f_g^{\gamma}\sim
{x^{-1.4}}$
at the scale $Q^2=p_T^2=4GeV$.
The
cross sections obtained using DG photon
structure function are more realistic, since the extrapolation
of DO parametrization to small $x$ region give unphysically
singular behavior.
For this reason, the
cross sections obtained using DO function should be treated only
as
an illustration of the strong dependence of the cross section
to a different choice of the photon structure function.
\vskip 3.58 true in
\vbox{
\tenrm\baselineskip=12pt
\noindent
Figure 1:
Total inelastic cross section ($\sigma_{soft}+\sigma_{JET}$)
predictions for HERA energies$^1$, compared to the recent ZEUS and H1
measurements$^2$.  The jet part includes contributions from
direct processes.  Shown separately are contributions
of direct processes, added to the constant soft part (curve $5$).
}
\vskip 0.3 true cm
\par
In Fig. 1 we also present the results for the cross section when only
soft and direct part are included, indicating its very weak energy
dependence.  The rise of the total cross section is thus
mostly driven by the
``anomalous'' (hadronic) part of the cross section.  We note that
HERA measurement has some resolving power to distinguish
between different sets of photon structure function and therefore
determine presently unknown low $x$ behavior of its gluon part.
For example, the cross sections
obtained using DO structure functions are already excluded by
HERA data, while
theoretical result obtained using DG structure function and
$p_{\perp,\rm min}=2GeV$
is consistent with the data (see Fig. 1).
However,
one should keep in mind that all the
theoretical predictions presented in Fig. 1 do not take into
account multiple scatterings for which one needs to use
proper eikonal treatment of high-energy scattering process.
The eikonalization procedure results (effectively)
in reducing the
cross section at HERA energies by
$10\%$ for $p_
{\perp,\rm min}=2GeV$ and by about $30\%$ for
$p_{\perp,\rm min}=1.41GeV$$^{12}$.
\par
We have also calculated the total
jet cross sections at
Fermilab and HERA energies for jet $p_T$ triggers of
$3,4,5GeV$ and $5,10,15GeV$ respectively$^1$.  The energy
dependence of the total jet cross section is much steeper than of the
direct part of the jet cross section only.
This can be seen in Figs. 4 and
5 in Ref. 1.
Again, jet measurements at HERA energies can
distinguish between different photon structure functions,
but
in this case one is probing the photon structure function
at slightly higher values of $x$ than in the case of the
total
cross section.
\par
\bigskip
\noindent
\line{\elevenbf 3. The Hadronic Photon and the ``Muon Puzzle'' \hfil}
\vglue 0.4cm
The ultra-high energy photoproduction
cross sections play
an
important role in
understanding recently observed anomalous muon
content in cosmic ray air-showers associated with
astrophysical
``point'' sources (such as Cygnus X-3, Hercules X-1 and Crab
Nebula)$^3$.  The number of muons
observed is
comparable with what one would expect in a hadronic shower, but
the fact that primary particle has to be long-lived and
neutral,
makes photon the only candidate in the Standard Model.
Conventionally, one would expect that photon produces electromagnetic
cascade and therefore muon poor.  However, if the photonuclear
cross section at very high energies becomes comparable with pair
production and bremsstrahlung
cross section ($\sigma_{\gamma \rightarrow
e^+e^-}\sim 500mb$) the muon content in a photon initiated
shower will be affected.  The hadronic character of the photon
enchances the photonuclear cross section at very high energies.
We have calculated the
total inelastic photon-air cross
sections in a QCD-based diffractive model, which takes into
account unitarity constraints necessary at
ultra-high energies$^{13,12}$.
\par
Our results for the photon-proton
and photon-air cross
sections at energies $10GeV\leq \sqrt s\leq 10^4GeV$ are
presented in Fig. 6 in Ref. 12.
We find that $\sigma_{\gamma - p}$ at $\sqrt s\geq 10^3 GeV$ is
about two times
larger than the conventional value, large enough to
be interesting, but much too small to account for the reported
muon anomalies in photon-initiated showers.
\bigskip
\line{\elevenbf 4. Conclusion \hfil}
\vglue 0.4cm
We have shown how measurement of the total
photoproduction cross section
at
HERA energies provides valuable information about
the hadronic
character of the photon.
In particular we emphasized how
measurements of $\sigma_{\gamma p}$ at HERA energies
can impose strong constraint on the value
of the theoretical jet momentum cutoff and, more importantly,
determine the small $x$ behavior
of the photon structure function and its gluon content.
With the theoretical uncertainties being reduced, we have
extrapolated our predictions for the photonuclear cross section to
ultra-high energies relevant for cosmic ray experiments.
The new results on
$\gamma$-air interactions make it quite clear that the hadronic
interactions of the photon cannot explain the reported muon
anomalies in cosmic ray air-showers, if the anomalies in fact exist.
Furthermore, future cosmic
ray experiments might be able to
put the ``muon puzzle'' observations on
firmer grounds and to provide valuable input to
particle physics at ultra-high energies, currently far beyond
the range of accelerator
experiments.
\bigskip
\line{\elevenbf 5. Acknowledgements \hfil}
\vglue 0.4cm
The work presented here was done in collaboration with
R. Gandhi whom I would like to thank for many useful
discussions.  This work was
supported in
part by the United States Department
of Energy Grants Nos. DE-FG02-85ER40213 and DE-FG03-93ER40792.
\bigskip
\line{\elevenbf 6. References \hfil}
\vglue 0.4cm
\item{1.} R. Gandhi and I. Sarcevic, {\elevenit Phys. Rev.}
{\elevenbf D44}, 10 (1991).
\item{2.}  M. Derick {\elevenit{et al.}},
ZEUS Collaboration, {\elevenit Phys. Lett.} {\elevenbf B293}, 465 (1992);
T. Ahmed {\elevenit et al.},
H1 Collaboration, {\elevenit Phys. Lett.} {\elevenbf B299}, 374 (1993).
\item{3.} M. Samonski and W. Stamm, {\elevenit Ap. J.}~{\elevenbf
{L17}},
268 (1983);
B. L. Dingus {\elevenit et. al.}, {\elevenit Phys. Rev. Lett}~
{\elevenbf{61}}, 1906 (1988);
Sinha {\elevenit et al.}, Tata Institute preprint, OG 4.6-23;
T. C. Weekes, {\elevenit Phys. Rep.}~{\elevenbf{160}}, 1 (1988),
and reference
therein.
\item{4.} D. O. Caldwell {\elevenit et al.},
{\elevenit Phys. Rev. Lett.} {\elevenbf 25},
609 (1970);
{\elevenit Phys. Rev. Lett.}
{\elevenbf 40},
1222 (1978);
H. Meyer {\elevenit et al.}, {\elevenit Phys. Lett.} {\elevenbf 33B},
189 (1970);
T. A. Armstrong {\elevenit et al.}, {\elevenit Phys. Rev.}
{\elevenbf D5}, 1640 (1972);
S. Michalowski {\elevenit et al.}, {\elevenit Phys. Rev. Lett.}
{\elevenbf 39}, 733 (1977).
\item{5.} N. Amos {\elevenit et al.},
{\elevenit Nucl. Phys.} {\elevenbf B262} 689 (1985); R. Castaldi and
G. Sanguinetti, {\elevenit Ann. Rev. Nucl. Part. Sci.}
{\elevenbf 35}, 351 (1985);
 M. Bozzo {\elevenit et al.},
{\elevenit Phys. Lett.} {\elevenbf 147B}, 392 (1984);
G. Alner {\elevenit et al.},
 {\elevenit Z. Phys.} {\elevenbf C32}, 156 (1986);
T. Hara {\elevenit et al.}, {\elevenit Phys. Rev Lett.}
{\elevenbf 50}, 2058
 (1983); R. M. Baltrusaitis
 {\elevenit et al.}, {\elevenit Phys. Rev. Lett.}
 {\elevenbf 52}, 1380 (1984).
\item{6.} D. Cline, F. Halzen and J. Luthe, {\elevenit Phys. Rev. Lett.}
{\elevenbf 31}, 491 (1973);
T. K. Gaisser and F. Halzen, {\elevenit Phys. Rev. Lett.}
{\elevenbf 54}, 1754 (1985);
L. Durand and H. Pi, {\elevenit Phys. Rev. Lett.} {\elevenbf 58},
303 (1987).
\item{7.} C. Albajar {\elevenit{et al.}}, UA1 Collaboration,
{\elevenit Nucl. Phys.} {\elevenbf B309}, 405 (1988).
\item{8.}  T. Ahmed {\elevenit et al.}, H1 Collaboration,
{\elevenit Phys. Lett.} {\elevenbf B297}, 205 (1992);
M. Derick {\elevenit et al.},
ZEUS Collaboration,
{\elevenit Phys. Lett.} {\elevenbf B297}, 404 (1992).
\item{9.} M. Drees and K. Grassie, {\elevenit Z. Phys.}
{\elevenbf C28}, 451 (1985);
D. Duke and J. Owens, {\elevenit Phys. Rev.}
{\elevenbf D26}, 1600 (1982); H. Abramowicz
{\elevenit et al.}, {\elevenit Phys. Lett.} {\elevenbf B269}, 458 (1991).
\item{10.} I. Sarcevic, S. D. Ellis and P. Carruthers,
{\elevenit Phys. Rev.} {\elevenbf D40}, 1472 (1989).
\item{11.} E. Eichten, I. Hinchliffe, K. Lane and C. Quigg,
{\elevenit Rev. Mod. Phys.} {\elevenbf 56}, 579 (1984).
\item{12.} L. Durand, K. Honjo, R. Gandhi, H. Pi and I. Sarcevic,
{\elevenit Phys. Rev.} {\elevenbf D47}, 4815 (1993);
{\elevenit Phys. Rev.} {\elevenbf D48}, 1048 (1993); L. Durand, these
Proceedings.
\item{13.} R. Gandhi, I. Sarcevic, A. Burrows, L. Durand and H. Pi,
{\elevenit Phys. Rev.} {\elevenbf D42}, 263 (1990).
\eject
\bye
\end